\documentclass{PoS}
\usepackage{color}

\title{Excited state systematics in extracting nucleon electromagnetic form factors}

\ShortTitle{Excited state systematics in extracting nucleon electromagnetic form factors}

\author{S.~Capitani$^{1,2}$, M.~Della~Morte$^{1,2}$, G.~von~Hippel$^{1}$, B.~J\"ager$^{1,2}$, 
	B.~Knippschild$^{1}$, H.B.~Meyer$^{1,2}$,\speaker{T.D.~Rae}\thanks{Supported by DFG grant HA4470/3-1}$~^{1}$, H.~Wittig$^{1,2}$\\
       
        $^{1}$ PRISMA Cluster of Excellence and Institut f\"ur Kernphysik, Becher-Weg 45, University of
	Mainz, D-55099 Mainz, Germany\\
	$^{2}$ Helmholtz Institute Mainz, University of Mainz, D-55099 Mainz, Germany\\
 	E-mail: \email{thrae@uni-mainz.de}}

\abstract{\vspace{-9.5cm} \phantom{a} \hfill HIM-2012-5 \newline \phantom{a}
\vspace{8.5cm} \newline 
We present updated preliminary results for the nucleon electromagnetic form factors for non-perturbatively $\mathcal{O}(a)$  
improved Wilson fermions in $N_f=2$ QCD measured on the CLS ensembles. The use of the summed 
operator insertion method allows us to suppress the influence of excited states 
in our measurements. A study of the effect that excited state contaminations have on the $Q^2$ dependence 
of the extracted nucleon form factors may then be made through comparisons of the summation method to 
standard plateau fits, as well as to excited state fits.
\newline

\begin{flushright}
\includegraphics[width=0.28\linewidth]{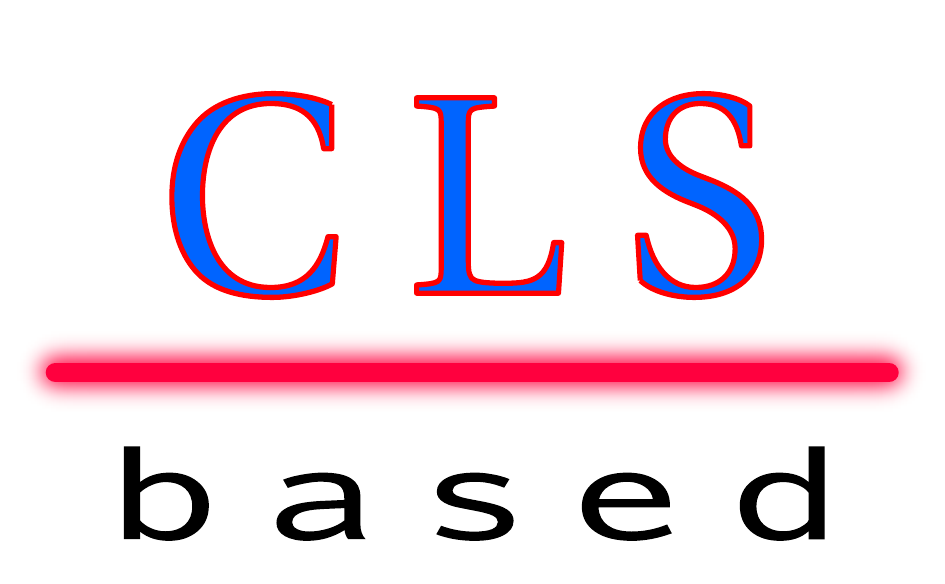}
\end{flushright}
}

\FullConference{The 30 International Symposium on Lattice Field Theory - Lattice 2012,\\
		June 24-29, 2012\\
		Cairns, Australia}

\begin{document}

\section{Introduction\label{seci}}
Baryon form factors are central observables of hadronic physics and provide details of the object's distribution of charge and magnetisation.
Currently, lattice simulations fall short of the accuracy achieved by experiment. Furthermore, the simulations of the nucleon electromagnetic form factor fail to 
reproduce experimental results \cite{Alexandrou:2010cm,Renner:2010ks}. It is therefore important to ensure that systematic effects are under control in lattice simulations. To study one such effect, we employ three 
separate methods (described in section~\ref{seciii}) that account for excited state contributions with varying degrees of rigour to check the control of systematic errors in the extraction of the form factors. 
This work provides an update on results previously presented in \cite{Capitani:2010sg} and follows the methodology used in a recent study of the nucleon's axial form factor \cite{Capitani:2012gj}. Similar methods have been used in \cite{Green:2012ud}.
Our simulations use non-perturbatively $\mathcal{O}(a)$ improved 
Wilson fermions in $N_f=2$ QCD, measured on the CLS ensembles. Table~\ref{ensembles} provides details of the lattice ensembles. 

\begin{table}
\centering
\begin{tabular}{cccccc}
\hline
$\beta$ & $a[\textrm{fm}]$ & $L/a$ & $L[\textrm{fm}]$ & $m_\pi[\textrm{MeV}]$ & $\textrm{no. meas}$  \\
\hline
\hline
$5.2$ & $0.079$ & $32$ & $2.5$ & $312,~363,$ & $1696,~796,$ \\
$ $   & $ $     & $ $  & $   $ & $473,~603$ & $1060,~576 $ \\
\hline
$5.3$ & $0.063$ & $32$ & $2.0$ & $451,~606,~649,$ & $1344,~1296,~278,$ \\
$ $   & $ $     & $48$ & $3.0$ & $277,~324$  & $1000,~796$ \\
\hline
$5.5$ & $0.050$ & $48$ & $2.4$ & $430,~536 $ & $600,~600$ \\
\hline
\end{tabular}
\caption{\label{ensembles} Details of the lattice ensembles used in this study, showing $\beta$-values, lattice spacing $a$, lattice extent $L$ (where $T=2L$),  
 pion mass $m_\pi$ and total number of measurements.}
\end{table}

The matrix element of a nucleon interacting with an electromagnetic current $V^\mu=\overline{\psi}(x)\gamma^\mu \psi(x)$ 
may be decomposed into the Dirac and Pauli form factors $F_1$ and $F_2$:
\begin{equation}
\langle N(p^\prime,s^\prime)|V_\mu|N(p,s)\rangle=\bar{u}(p^\prime,s^\prime)\left[\gamma_\mu F_1(Q^2)+i\frac{\sigma_{\mu\nu} q_\nu}{2m_N}F_2(Q^2)\right]u(p,s),
\end{equation}
where $u(p,s)$ is a Dirac spinor with spin $s$, and momentum $p$, $\gamma_\mu$ is the Dirac matrix, and $\sigma_{\mu\nu}=\frac{1}{2i}[\gamma_\mu,\gamma_\nu]$.
Also $Q^2=-(E_{p^\prime}-E_p)^2+\vec{q}^2$ where $\vec{q}=\vec{p}^\prime-\vec{p}$.
These form factors are related to the Sachs form factors, $G_E$ and $G_M$, 
\begin{equation}
G_E(Q^2)=F_1(Q^2)-\frac{Q^2}{4m_N^2}F_2(Q^2), \qquad G_M(Q^2)=F_1(Q^2)+F_2(Q^2),
\end{equation}
that are measured in scattering experiments via the differential cross section described by the Rosenbluth formula.
The form factors may be Taylor expanded in the momentum transfer $Q^2$
\begin{equation}
G_X(Q^2)=G_X(0)\left(1+\frac{1}{6}\langle r^2 \rangle Q^2+\mathcal{O}(Q^4)\right),
\end{equation}
from which the charge radii of the nucleon may be determined:
\begin{equation}
\langle r_X^2 \rangle = \frac{6}{G_X(Q^2)}\frac{\partial G_X(Q^2)}{\partial Q^2}\Bigg|_{Q=0},
\end{equation}					
where $X=E,~M$. For the conserved current, $G_E(0)=1$ and $G_M(0)=\mu$, where $\mu$ measures the magnetic moment in nuclear magneton units $e/(2m_N)$.

\section{Lattice formulation\label{secii}}

The calculation of the form factors requires a ratio of correlation functions, for which we use (for the case $\vec{p}^\prime=0$)
\begin{eqnarray}
R_{\gamma_\mu}(\vec{q},t,t_s)&=&\frac{C_{3,\gamma_\mu}(\vec{q},t,t_s)}{C_2(\vec{0},t_s)}\sqrt{\frac{C_2(\vec{q},t_s-t)C_2(\vec{0},t)C_2(\vec{0},t_s)}{C_2(\vec{0},t_s-t)C_2(\vec{q},t)C_2(\vec{q},t_s)}},
\label{Ratio}\\
C_2(\vec{p},t)&=&\sum_{\vec{x}}\langle\Gamma_{\alpha^\prime\alpha}J_\alpha(x)\overline{J}_{\alpha^\prime}(0)\rangle e^{-i\vec{p}.\vec{x}},\\
C_{3,\gamma_\mu}(\vec{q},t,t_s)&=&\sum_{\vec{x},\vec{y}}\langle\Gamma_{\alpha^\prime\alpha}J_\alpha(\vec{x},t_s)\mathcal{O}_{\gamma_\mu}(\vec{y},t)\overline{J}_{\alpha^\prime}(0)\rangle e^{-i\vec{q}.\vec{y}}.
\end{eqnarray}
This ratio was found to be the most effective in \cite{Alexandrou:2008rp}. $C_2(\vec{p},t)$ and $C_{3,\gamma_\mu}(\vec{q},t,t_s)$
are two- and three-point functions respectively. $J_\alpha(x)$ is a suitably chosen interpolating operator with the correct quantum numbers to create a nucleon,
and $\Gamma_{\alpha\alpha^\prime}$ is a projection matrix used to give the interpolating fields the correct parity. We choose to polarise the nucleon in the $z$-direction, 
for which $\Gamma=\frac{1}{2}(1+\gamma_0)(1+i\gamma_5\gamma_3)$. This choice allows us to extract both $G_E$ and $G_M$,
whereas for an unpolarised 
nucleon, $\Gamma=\frac{1}{2}(1+\gamma_0)$, one may only determine $G_E$. Further to this, for improved statistics we average over polarisation in 
the positive and negative $z$-direction as well as over the nucleon and anti-nucleon state. 

Due to the insertion of the operator at time $t$, the three point function is a more computationally demanding object. We use the `fixed sink method' for its calculation, which fixes the final and 
initial states, but allows both the operator and momentum transfer to be chosen without the need for additional inversions \cite{Martinelli:1988rr}. For this study, we consider both
the local and conserved vector current for the operator $\mathcal{O}_\mu$. The latter is defined as
\begin{eqnarray}
\mathcal{O}^\textrm{con}_\mu(x)=\frac{1}{2}\left(\overline{\psi}(x+a\hat{\mu})(1+\gamma_\mu)U^\dagger_\mu(x)\psi(x)-\overline{\psi}(x)(1-\gamma_\mu)U_\mu(x)\psi(x+a\hat{\mu})\right)
\end{eqnarray}
where $\psi=u,d$. In principle, we are able to determine the form factors for the proton, neutron, iso-scalar and iso-vector depending on the linear combination of contributions 
from the quark correlation functions. We concentrate on the iso-vector,
which has the advantage that quark-disconnected diagrams cancel.

To improve the overlap of the interpolating operators with the nucleon, we used Gaussian smearing at both the source and sink supplemented by HYP smeared links. Whilst this is a great improvement over point sources, the extraction of plateaus at non-zero momenta can still prove difficult. A poster presented at this conference addresses this problem for two-point functions, whereby
using a generalisation of Gaussian smearing to form an anisotropic wavefunction results in a reduction of the noise-to-signal ratio at non-zero momenta for the pion and nucleon \cite{Poster,DellaMorte:2012xc}. We are currently extending this study to include three-point correlation functions with the aim of applying the technique to the extraction of the vector form factors. 

$G_E$ and $G_M$ may be extracted from eq.~(\ref{Ratio}) at large time arguments: 
\begin{equation}
R_{\gamma_0}(\vec{q})=\sqrt{\frac{M+E}{2E}}G_E(Q^2),\qquad R_{\gamma_i}(\vec{q})=\epsilon_{ij}p_j\sqrt{\frac{1}{2E(E+M)}}G_M(Q^2)\quad i=1,2.
\end{equation}

\section{Systematics of extraction\label{seciii}}

\begin{figure}
\centering
\includegraphics[width=0.49\linewidth]{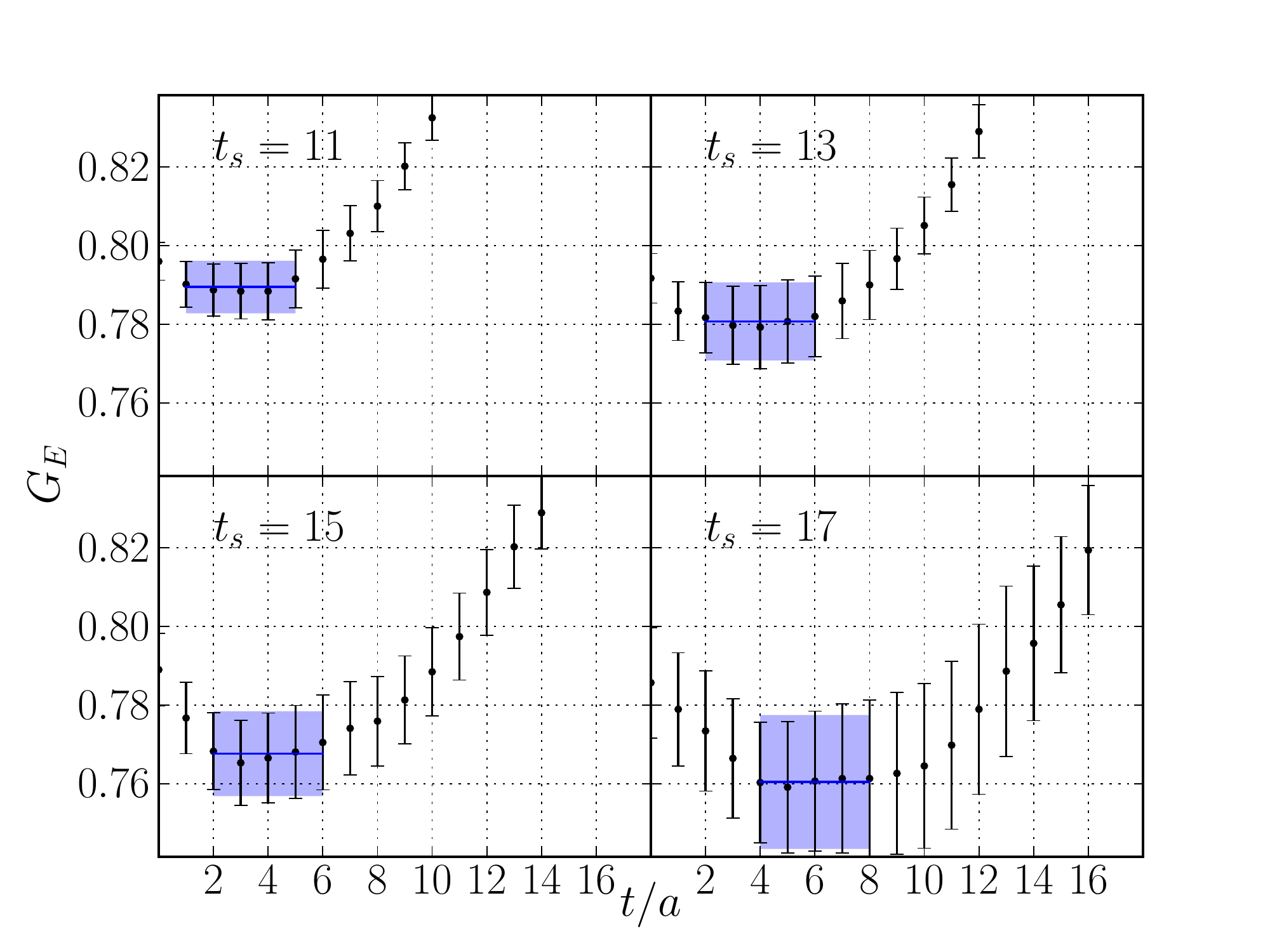}
\includegraphics[width=0.49\linewidth]{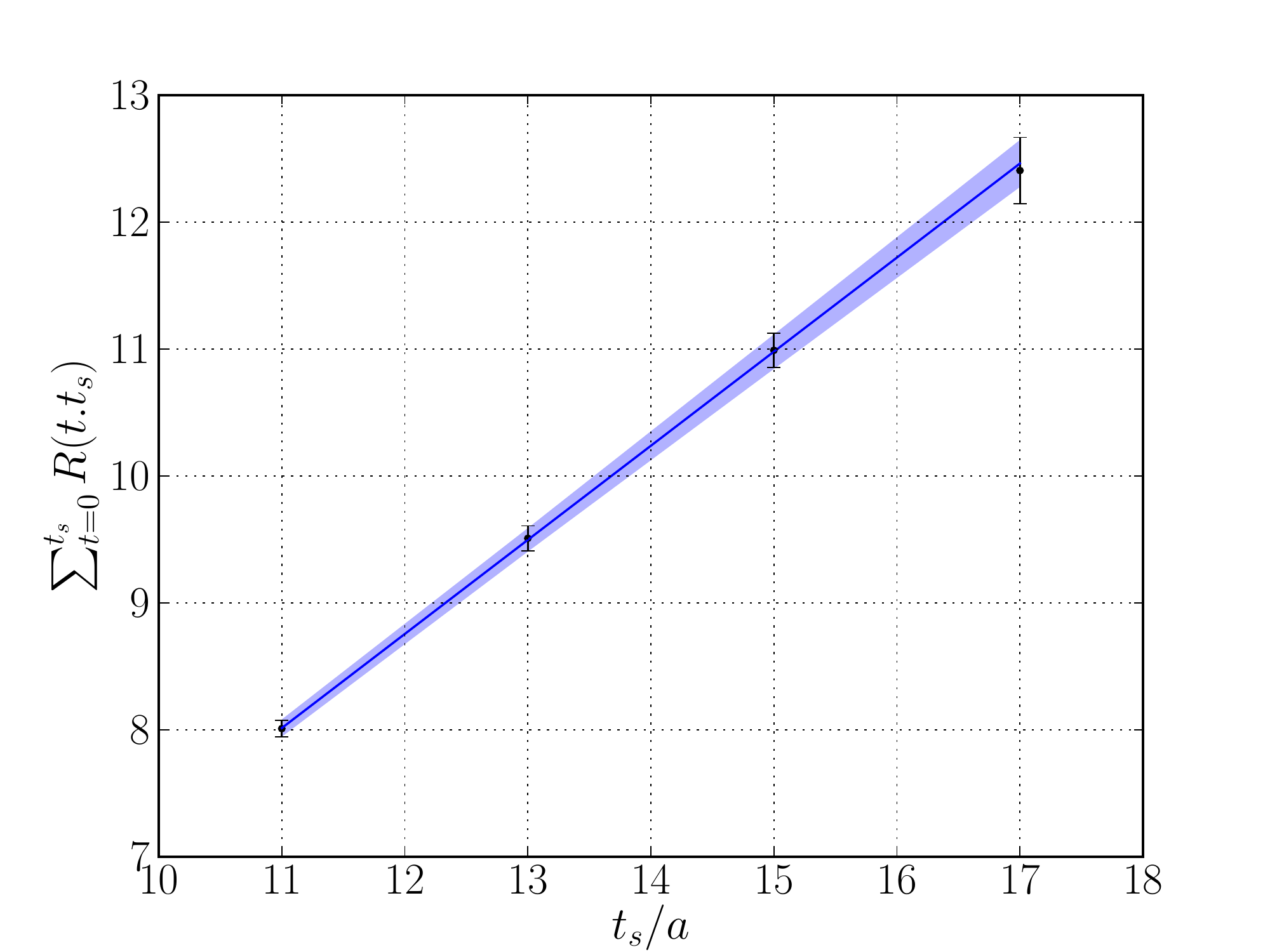}
\caption{\label{tsdep}\small Left panel: Plateau fits to $G_E$ for several $t_s$ for the smallest non-zero momentum transfer, $Q^2$. Excited state contamination effects for the smallest $t_s$ are clearly visible (the asymmetric distribution of points is a remnant of the non-zero momentum transfer). Right panel: Summation method for $G_E$. Both panels are shown for our lightest ensemble $(m_\pi=277~\textrm{MeV})$.}
\end{figure}

Correlation functions must have reached their asymptotic behaviour for a reliable and unbiased determination of the form factors.
However, we observe exponentially decaying excited states from both the source and sink.
Therefore, simple plateau fits (left panel fig.~\ref{tsdep}) show a trend of higher values for small source-sink separations, i.e. for decreasing $t_s$. 
To control systematics, it is important to take these excited states into account.
Contributions to the ratio from the ground and excited states may be factorised
\begin{eqnarray}
R(\vec{q},t,t_s)=R^0(\vec{q},t,t_s)\Big(1+\mathcal{O}\big(e^{-\Delta t}\big)+\mathcal{O}\big(e^{-\Delta^\prime(t_s-t)}\big)\Big)
\end{eqnarray}
where $\Delta$ and $\Delta^\prime$ are the energy gaps of the initial and final nucleons respectively. With the assumption $\Delta=\Delta^\prime=2m_\pi$\footnote{This is not strictly true when there is a momentum transfer, however we find this to be a small effect as the data is well described by
eq.~(\ref{exeq}). Further to this, the results agree with the summation method, which does not require an assumption for the energy gap.} we take the excited states into account using a fit to 
\begin{equation}
R(\vec{q},t,t_s)=G_{E,M}+b_1 e^{-\Delta t}+b_2 e^{-\Delta(t_s-t)}+b_3 e^{-\Delta t_s}.\label{exeq}
\end{equation}
An alternative without the need for the assumption $\Delta=\Delta^\prime=2m_\pi$ uses summed operator insertions \cite{Sum}:
\begin{equation}
S(t_s)=\sum_{t=0}^{t_s}R(\vec{q},t,t_s)\rightarrow c(\Delta,\Delta^\prime)+t_s\left(G_{E,M}+\mathcal{O}\big(e^{-\Delta t_s}\big)+\mathcal{O}\big(e^{-\Delta^\prime t_s}\big) \right).
\end{equation}
This allows the form factors to be extracted from the slope, from computing $S(t_s)$ for several $t_s$ (right panel fig.~\ref{tsdep}). The excited states should be more suppressed for this method than for a fit to eq.~(\ref{exeq}), because $t_s>t,(t_s-t)$.

\section{Form factor $Q^2$ dependence and chiral behaviour\label{seciiii}}

\begin{figure}
\centering
	\includegraphics[width=0.80\linewidth]{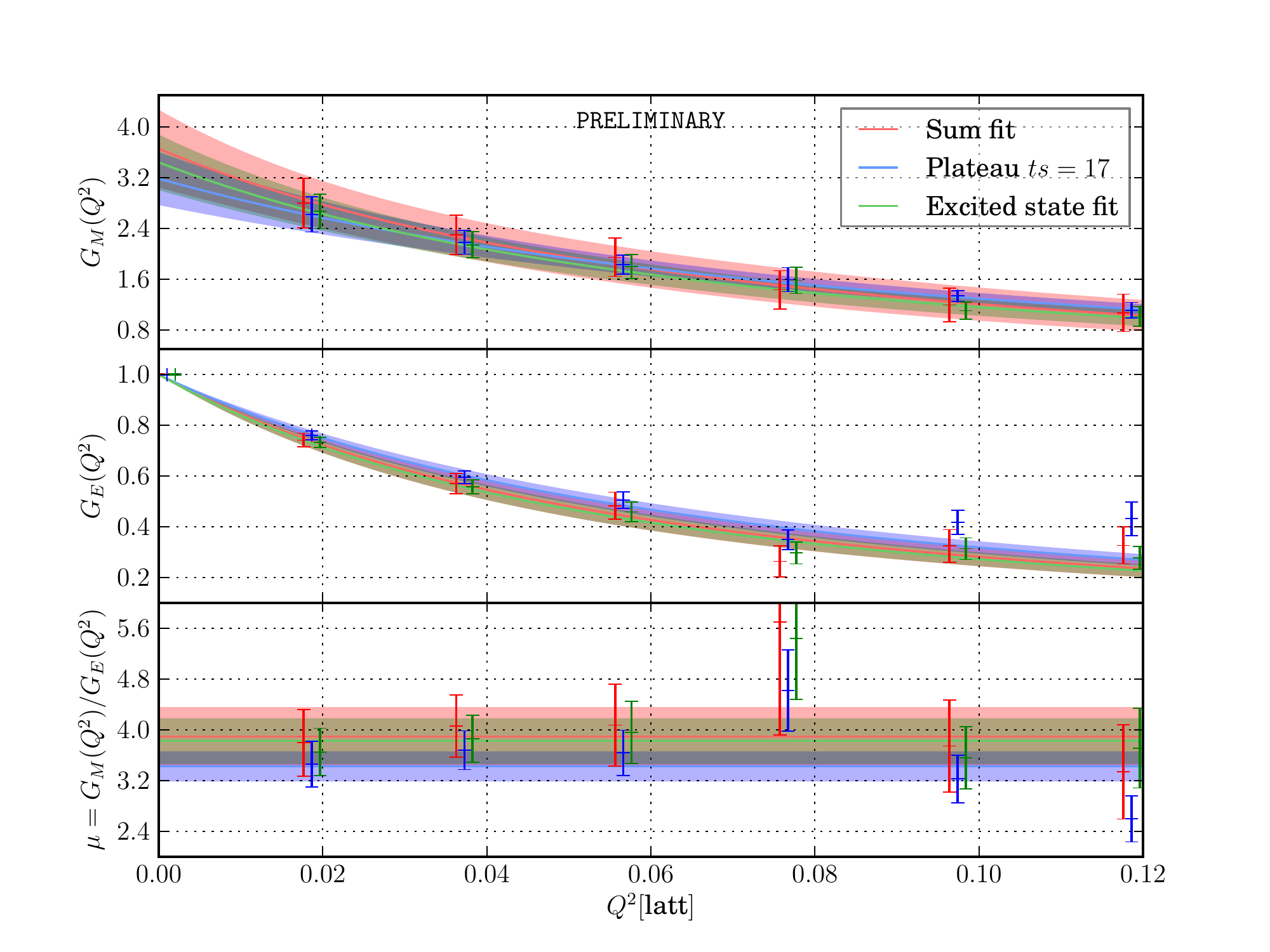}
	\caption{\label{qdep} Results for our most chiral ($m_\pi=277~\textrm{MeV}$) lattice. The top two panels show the $Q^2$ dependence for $G_E$ and $G_M$. The bottom panel shows a determination of the magnetic moment, $\mu$. This is shown for the three extraction methods.}
\end{figure}

For the remaining discussion we concentrate on the conserved current as this removes the need for any renormalisation of the lattice operators; however, we note that a comparison between the local and conserved
current provides a check of the renormalisation factor, which we find to be in agreement with other work (e.g.~\cite{Della Morte:2005rd}).
In order to model the $Q^2$ dependence of the form factors (shown in fig.~\ref{qdep} for our most chiral analysed ensemble) a dipole ansatz is adopted
\begin{equation}
G_{E,M}(Q^2)=\frac{G_{E,M}(0)}{\left(1+\frac{Q^2}{M_{E,M}^2}\right)^2},
\end{equation}
from which the radii is extracted. We may also obtain the magnetic moment $\mu$ from $G_M(0)$ and also from the ratio
\begin{equation}
\mu=\lim_{Q^2\rightarrow0}\frac{G_M(Q^2)}{G_E(Q^2)},
\end{equation}
based upon the phenomenological observation that the form factors $G_E$ and $G_M$ have very similar radii. For our determination, the two methods are in good agreement (fig.~\ref{qdep}).

This calculation was performed on the ensembles listed in table~\ref{ensembles} to check both finite volume and discretisation effects. 
The lattice ensembles cover a range of pion masses (277 to 649~MeV). In order to perform an extrapolation in the pion mass to the physical point, we model the chiral behaviour of the radii using a HB$\chi$PT inspired fit \cite{Khan:2006de},
\begin{eqnarray}
\langle r_1^2 \rangle = c_1 + c_2 \log(m_\pi^2).
\end{eqnarray}
We are currently investigating the effect of other fit forms on the result, including linear fits and covariant B$\chi$PT fits as well as applying cuts to the pion mass range \cite{Collins:2011mk,Alexandrou:2011db,Bratt:2010jn,Yamazaki:2009zq}, so as to have a comprehensive picture of the systematic effects. We also apply this to $\langle r_2^2 \rangle$ and $\kappa=\mu-1$.

\begin{figure}
\centering
	\includegraphics[width=0.72\linewidth]{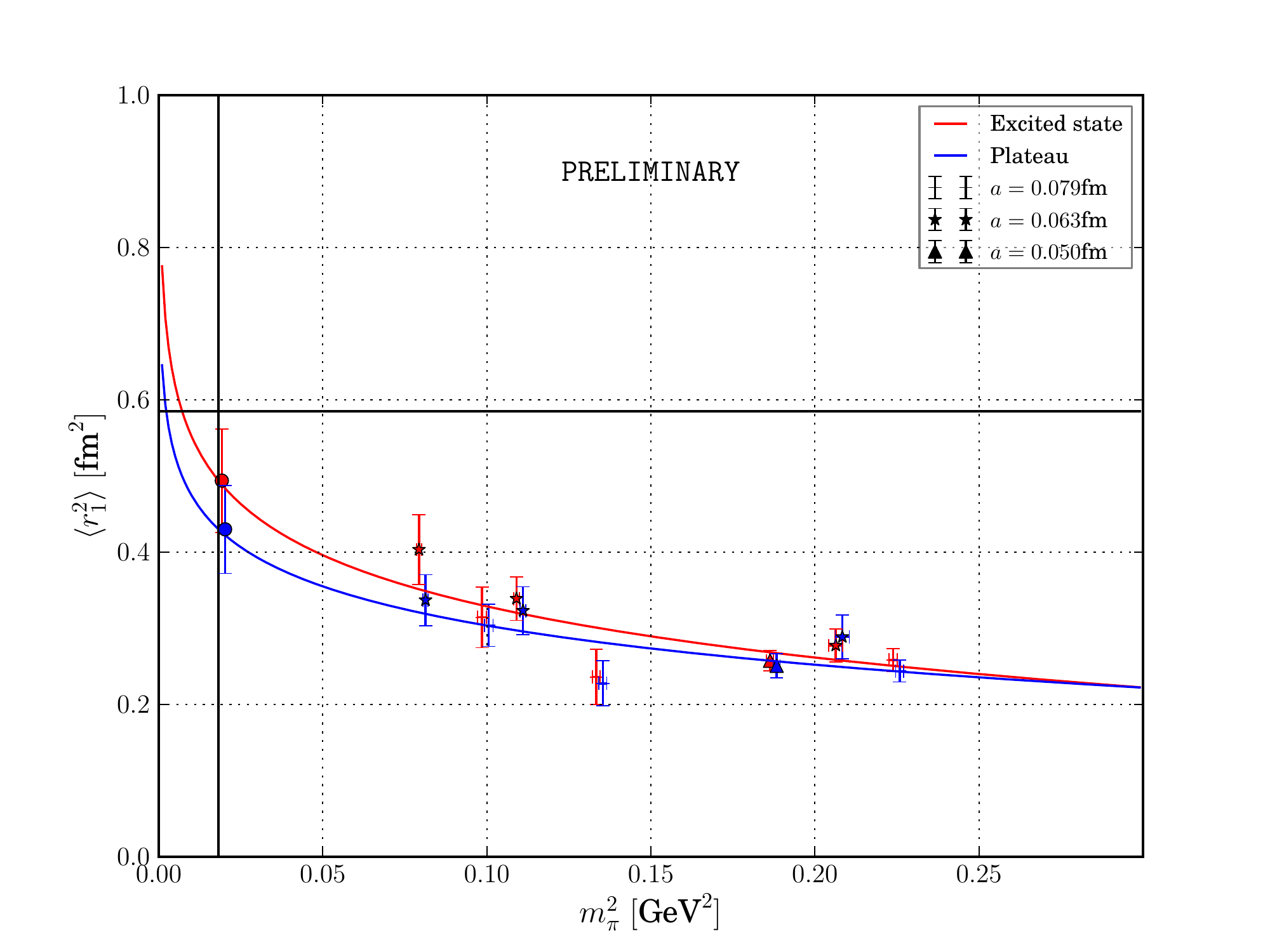}
	\caption{\label{chi} Chiral extrapolation of the Dirac radius to the physical point (vertical black line) using $\langle r_1^2\rangle=c_1+c_2\log(m_\pi^2)$ using the largest $t_s$ plateau-fit (blue) and excited state fit (red) for the extraction. The horizontal black line shows the experimental result. The different symbols indicate the lattice spacing (see legend) and the leftmost two points show our extrapolated values.}
\end{figure}

Thus far we have sketched the methodology commonly employed to extract form factors on the lattice. 
Using this, we study the systematics of the extraction by separately employing the three analyses 
described above: a plateau fit (for the largest $t_s$), a simultaneous excited state fit to $t$ and $t_s$
and the summation method. A comparison of the three methods in fig.~\ref{qdep} shows that whilst all three methods agree within statistical errors, a systematic trend for better agreement between the two methods that account for excited state effects is observed. This is echoed in all of our ensembles. An earlier study by our group on $g_A$ \cite{Capitani:2012gj} demonstrates a clearer systematic dependence. However, we note that $g_A$ is a somewhat cleaner quantity to extract on the lattice (i.e. $Q^2=0$). The electromagnetic form factor analysis requires an improvement in the statistical errors and the addition of smaller pion masses to further check the systematic behaviour and as such the results in this proceeding should be considered as preliminary. Fig.~\ref{chi} compares the determination of the Dirac radius $\langle r_1^2\rangle$ from an excited state fit with the plateau fit (the summation method agrees well with the excited state fit but with increased statistical errors). We observe an increase in the central value for all but one ensemble and also that the difference increases as the pion mass decreases (as one would expect, excited states should contribute more for more chiral $m_\pi$). This results in a larger radius that is closer to the experimental result. If we restrict the fit to the four most chiral ensembles and perform a straight line fit, a result compatible with experiment is achieved for the excited state method for $\langle r_1^2\rangle$. We have also looked at $\kappa$ and $\langle r_2^2\rangle$, for which we see a similar behaviour. These are however all at the preliminary stage.

\section{Conclusions and outlook\label{seciiiii}}

We have presented preliminary results for the nucleon vector form factors with a particular emphasis on the potential for systematic errors from excited state contaminations.
We account for this through a comparison of three extraction techniques which account for excited states with varying degrees of rigour.
Whilst we are, so far, unable to see any effects outside of errors between the three methods, we do see an improved agreement between the two methods that account for excited state effects in both the $Q^2$ behaviour and in the chiral extrapolations of the form factors, and consider this to be an indication of a systematic effect. Furthermore, there is a trend for a steeper gradient to be extracted from these results and thus such a systematic could lead to better agreement between lattice results and experiment.
The results indicate the importance of excited states and that they should be a consideration in studies of potential systematic effects.

The large statistical errors for all methods highlight the need for greater statistics and for more chiral points to be added, where excited states should, in principle, have a greater effect. Monte Carlo ensembles exist for more chiral points (the lightest available is approximately 200~MeV), but are yet to be analysed.

\end{document}